\documentclass[%
 reprint,amsmath,amssymb,aps,nofootinbib
]{revtex4-2}

\usepackage{multirow}
\usepackage{graphicx}
\usepackage{bm}
\usepackage[dvipsnames]{xcolor}
\usepackage[
    colorlinks=true,
    linkcolor=blue,
    filecolor=blue,
    citecolor=blue,      
    urlcolor=blue,
    ]{hyperref}
\usepackage{array}

\newcolumntype{R}[1]{>{\raggedleft\let\newline\\\arraybackslash\hspace{0pt}}m{#1}}

\begin{document}

\title{Single-site DFT+DMFT for vanadium dioxide using bond-centered orbitals}

\author{Peter Mlkvik}\email{peter.mlkvik@mat.ethz.ch}
\author{Maximilian E. Merkel}
\author{Nicola A. Spaldin}
\author{Claude Ederer}
\affiliation{Materials Theory, Department of Materials, ETH Z\"{u}rich, Wolfgang-Pauli-Strasse 27, 8093 Z\"{u}rich, Switzerland}

\date{\today}

\begin{abstract}
We present a combined density-functional theory and single-site dynamical mean-field theory (DMFT) study of vanadium dioxide (VO$_2$) using an unconventional set of bond-centered orbitals as the basis of the correlated subspace. VO$_2$ is a prototypical material undergoing a metal-insulator transition (MIT), hosting both intriguing physical phenomena and the potential for industrial applications. With our choice of correlated subspace basis, we investigate the interplay of structural dimerization and electronic correlations in VO$_2$ in a computationally cheaper way compared to other state-of-the-art methods such as cluster DMFT. Our approach allows us to treat the rutile and M1 monoclinic VO$_2$ phases on an equal footing and to vary the dimerizing distortion continuously, exploring the energetics of the transition between the two phases. The choice of basis presented in this work hence offers a complementary view on the long-standing discussion of the MIT in VO$_2$ and suggests possible future extensions to other similar materials hosting molecular-orbital-like states.
\end{abstract}

\maketitle


\section{Introduction}

Vanadium dioxide (VO$_2$) is a prototypical example of a system undergoing a coupled structural and metal-insulator transition (MIT). This transition, occurring near room temperature at around $T_c$~=~340~K, has been heavily studied for decades, both due to the intriguing physics present in the system~\cite{Goodenough:1971, Zylbersztejn/Mott:1975, Qazilbash_et_al:2007, Wall_et_al:2018, Pouget:2021} and due to its potential for industrial uses \cite{Rini_et_al:2008,Yang/Ko/Ramanathan:2011,Wang_et_al:2016, Yi_et_al:2018}. Recently, VO$_2$ has also been identified as an obstructed atomic insulator~\cite{Xu_et_al:2021}. Because of the complex interplay of electronic and structural effects involving dimerization and potential correlation effects, multiple studies have focused specifically on methods for correctly simulating VO$_2$~\cite{Biermann_et_al:2005, Gatti_et_al:2007, Eyert:2011, Grau-Crespo/Wang/Schwingenschlogl:2012, Budai_et_al:2014, Brito_et_al:2016, Najera_et_al:2017, Grandi/Amaricci/Fabrizio:2020, Weber_et_al:2020}. Here, we present a method based on dynamical mean-field theory (DMFT) applied to a bond-centered orbital basis that is complementary to current state-of-the-art approaches and allows us to vary the structure continuously through the MIT between the two main VO$_2$ phases.

The first-order transition in VO$_2$ structurally transforms the system from a high-temperature high-symmetry rutile (R) phase into a low-temperature low-symmetry monoclinic (M1) phase~\cite{Eyert:2002}. At this transition, the nominally $d^1$ V$^{4+}$ cations undergo a dimerization along the $c$ direction of the R structure accompanied by the formation of a zig-zag pattern perpendicular to $c$. In the original picture proposed by Goodenough~\cite{Goodenough:1971}, the lowest-lying $a_{1g}$ orbitals, with lobes pointing along the dimerized chain, split into a bonding-antibonding pair, whilst the other two $t_{2g}$ orbitals, the so-called $e_{g}^{\pi}$ \footnote{We acknowledge that this is strictly not the correct symmetry label for the V site in both the R and M1 phases of VO$_2$. All three $t_{2g}$ levels are split and hence the two ``$e_g^\pi$'' levels are not degenerate in energy. However, we choose to continue the standard nomenclature as used in other references in the field.}, with lobes pointing away from the dimerized chain, are shifted upwards in energy and thus are depleted. However, this picture does not account for the fact that single-particle approaches such as density-functional theory (DFT) do not open a gap in the material~\cite{Eyert:2002} and do not capture the more exotic insulating monoclinic M2 phase which hosts separate dimers and zig-zagged chains~\cite{Pouget_et_al:1974, Rice/Launois/Pouget:1994, Quackenbush_et_al:2016}. 
This has led to a long-standing discussion in the literature about the character of the MIT in VO$_2$ with a pure Peierls-like distortion description thought to be insufficient and electronic correlations needed for a comprehensive description~\cite{Wentzcovitch/Schulz/Allen:1994, Rice/Launois/Pouget:1994}.

Many crucial developments in understanding the MIT in VO$_2$ have stemmed from the use of the DFT+DMFT methodology to account for strong local electron-electron interaction in the material. Early DFT+DMFT studies~\cite{Liebsch/Ishida/Bihlmayer:2005} established that although the R phase of VO$_2$ could be correctly described by effective single-particle approaches, a coherent treatment of the structural and electronic dimers is necessary to describe the M1 phase. The latter was first achieved in the seminal work by Biermann~{\it et al.}~\cite{Biermann_et_al:2005}, in which a single-site DFT+DMFT approach was used to study the R phase, while a DFT + two-site cluster DMFT approach encompassing the two dimerizing vanadium atoms was used for the M1 phase. In Ref.~\cite{Biermann_et_al:2005}, the authors described VO$_2$ as a ``dynamical-singlet insulator'' undergoing a correlation-assisted Peierls transition. 

In follow-up works~\cite{Tomczak/Biermann:2007, Tomczak/Aryasetiawan/Biermann:2008, Tomczak/Biermann:2009}, it was shown that including a frequency-independent interatomic self-energy between the vanadium atoms in a dimer captures the main behavior of the cluster DMFT approach, achieving remarkably similar results to the prior cluster DMFT study for the M1 phase. This avenue of research was further explored by Belozerov~{\it et al.}~\cite{Belozerov_et_al:2012}, who studied the M1 phase by employing an empirical intersite interaction using the DFT$+V$ approach plus a single-site DMFT on top, again yielding remarkably similar results to the prior cluster DMFT study. In contrast, a DFT + cluster DMFT study by Weber~{\it et al.}~\cite{Weber_et_al:2012}, using highly localized orbitals as the basis set, showed strong Mott effects suggesting that M1 VO$_2$ is a product of a Peierls-assisted orbital-selective Mott transition. More recently, another DFT + cluster DMFT study by Brito~{\it et al.}~\cite{Brito_et_al:2016, Brito_et_al:2017}, again using a localized atomic-like orbital basis, replicated some of the findings of Ref.~\cite{Biermann_et_al:2005}, but the authors concluded that VO$_2$ is primarily a Mott material. They also performed calculations of the M2 monoclinic phase and single-site DMFT for the R phase. Lastly, there have been multiple studies with other advanced methods such as $GW$, additionally highlighting the importance of interaction effects in VO$_2$ \cite{Gatti_et_al:2007, Weber_et_al:2020}.

\begin{figure}[!t]
	\centering
	\includegraphics[width=1\linewidth]{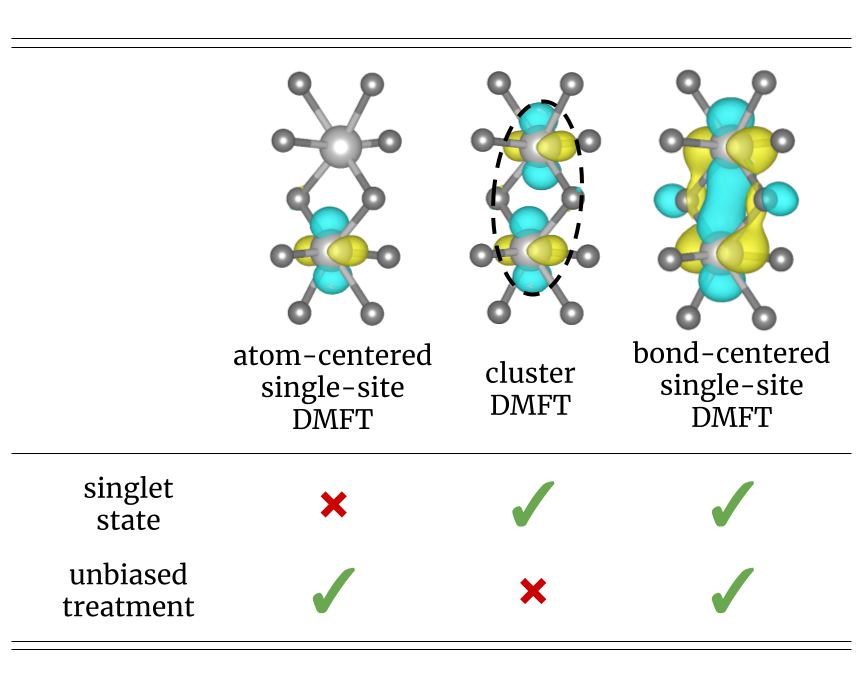}
	\caption{A schematic comparison of the differences between different DMFT-based methods for simulating VO$_2$: atom-centered single-site DMFT, two-site cluster DMFT, and -- the method presented in this work -- bond-centered single-site DMFT. We note that the bond-centered approach is unbiased because it considers orbitals for each V--V pair whereas in the cluster DMFT only short-bond V--V pairs in the M1 structure are treated as clusters.}
	\label{fig:basis_comp}
\end{figure}

In this work, we build on these previous works and, in particular, attempt to solve a key shortcoming in the literature -- the distinct treatment of the two main VO$_2$ phases in prior DFT+DMFT studies and the lack of an unbiased method for treating the R, M1, and in-between phases on an equal footing. To do so, we exploit the flexibility of the DFT+DMFT approach to choose an appropriate basis to describe interaction effects within the correlated subspace, motivated also by previous works on molecular-orbitals~\cite{Solovyev:2008, Kovacik/Ederer:2009, Kovacik_et_al:2012, Ferber_et_al:2014} or vacancy-centered orbitals~\cite{Souto-Casares/Spaldin/Ederer:2019, Souto-Casares/Spaldin/Ederer:2021}. We perform calculations based on bond-centered Wannier orbitals~(examples can be seen in Fig.~\ref{fig:bc-wfs}) that capture both intra- and inter-atomic effects when combined with DFT+DMFT at the single-site level. A bond-centered orbital set is a natural choice for VO$_2$ since it allows for the formation of a molecular-dimer-like singlet state that forms between dimerized V atoms in the monoclinic M1 phase~\cite{Hiroi:2015}. We construct bond-centered orbitals as the bonding combinations of atom-centered orbitals, directly capturing the bonding nature of the highest occupied valence band~\cite{Mlkvik/Ederer/Spaldin:2022}. Additionally, we construct our basis to offer the possibility to, in principle, condense an electronic singlet on either of the bonds in VO$_2$. This approach allows for a comprehensive treatment of VO$_2$ across the full range of distortion from R to M1 phase, without introducing a bias for dimerization by an {\it a priori} grouping of V atoms into dimer-forming pairs. At the same time, it is computationally less demanding than cluster DMFT. In Fig.~\ref{fig:basis_comp} we compare our bond-centered approach to the standard atom-centered single site and the previously used cluster DMFT methods, highlighting in particular the ability to describe singlet formation while providing an unbiased treatment of dimerization. 


\section{Methods}

\subsection{Construction of the bond-centered orbitals}

\begin{figure}
	\centering
	\includegraphics[width=1\linewidth]{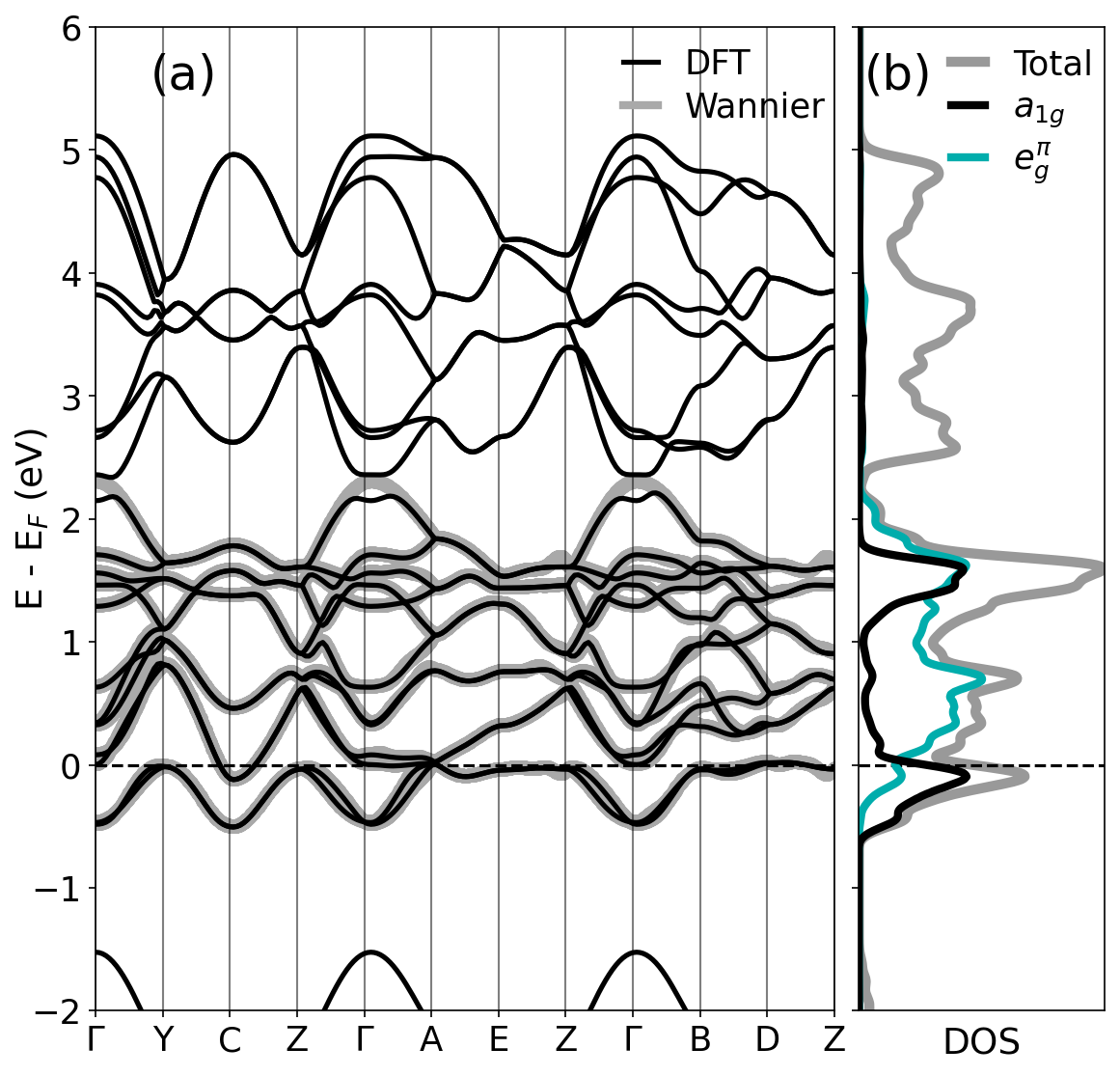}
	\caption{(a) DFT (Wannier) band structure of M1 VO$_2$ shown in black (grey). (b) Total DFT density of states (DOS) as well as orbital-projected DOS, with black (light-blue) lines indicating the $a_{1g}$ ($e_{g}^{\pi}$) component. The total DOS is shown in gray. The horizontal dashed line indicates the Fermi level $E_\text{F}$.}
	\label{fig:bs-plot}
\end{figure}

\begin{figure*}
	\centering
	\includegraphics[width=0.9\linewidth]{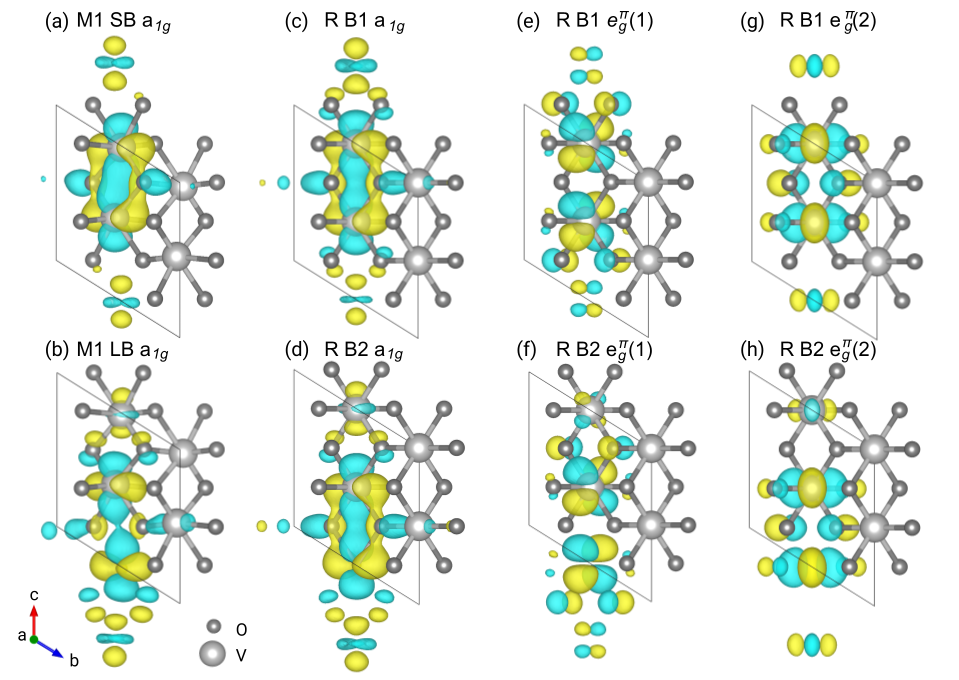}
	\caption{Bond-centered orbitals within the unit cells of the VO$_2$ structures. (a, b) M1 short-bond (SB) and long-bond (LB) $a_{1g}$ orbitals. (c, d) $a_{1g}$ orbitals for both bonds (B1 and B2) in the R structure, and (e-h) corresponding $e_{g}^{\pi}$ orbitals. V (O) atoms are shown in light (dark) gray and yellow (blue) isosurfaces at identical values for all plots show the positive (negative) phase of the orbital.}
	\label{fig:bc-wfs}
\end{figure*}

As outlined in the introduction, we perform DFT + single-site DMFT calculations for VO$_2$ using a basis of Wannier functions that are centered in between neighboring V atoms along the $c$ direction~(see Fig.~\ref{fig:bc-wfs}). We construct this basis from the V $t_{2g}$-dominated bands in the energy region between $-0.5$\,eV and $2.5$\,eV around the Fermi level, which are only slightly entangled with higher-lying V $e_g$-dominated bands (see the DFT band structure in Fig.~\ref{fig:bs-plot}).

We first construct a standard atom-centered basis using V-centered $t_{2g}$ orbitals as initial projections for the Wannier functions. We use this basis mainly as an intermediate step before constructing the bond-centered basis, but we also calculate the corresponding local interaction parameters using the constrained random phase approximation (cRPA) \cite{Aryasetiawan_et_al:2004, Miyake/Aryasetiawan:2008} for comparison with those of the bond-centered orbitals. 

To construct the bond-centered basis from these atom-centered orbitals, we then perform a pair-wise unitary transformation $U(\textbf{k})$ on the corresponding atom-centered Wannier functions in $\textbf{k}$-space, acting always on a pair of either $a_{1g}$ or $e_{g}^\pi$-type Wannier functions corresponding to the two neighboring V sites, 
\begin{equation} 
	\label{eq_rotmat}
	U(\textbf{k}) =  \frac{1}{\sqrt{2}}
	\begin{pmatrix}
		e^{i(\pi/4 - k_z c/4)} & e^{i(-\pi/4 + k_z c/4)}\\
		e^{i(-\pi/4 - 3 k_z c/4)} & e^{i(\pi/4 - k_z c/4)}\\
	\end{pmatrix}.
\end{equation}
Here, $c$ is the lattice parameter of the monoclinic structure along the dimerization directions, which is twice that of the standard rutile cell. 
In Appendix~\ref{app_rot-mat}, we show that the Wannier functions resulting from this transformation are indeed centered in the middle between consecutive V atoms along the $c$ direction.

We note that this transformation can either be performed by post-processing the final Wannier transformation matrices $U_{mn}(\mathbf{k})$ or by transforming the initial projection overlap matrices $A_{mn}(\mathbf{k})$ (see Ref.~\cite{Marzari_et_al:2012} for the definition of these matrices) without performing a subsequent spread minimization.

The use of bond-centered Wannier functions allows us to treat each V--V pair as a single-site impurity problem within the DFT+DMFT framework~\cite{Georges_et_al:1996, Held:2007, Beck_et_al:2022}. We point out that we construct ``bond-centered'' Wannier functions for each V--V pair along $c$, i.e., for both the dimerized short bond (SB) pair and the non-dimerized long-bond (LB) pair in the M1 structure (and for both equivalent V--V pairs in the R structure). Therefore, the atom-centered and bond-centered bases span exactly the same correlated subspace [marked by the thick grey lines in Fig.~\ref{fig:bs-plot}(a)], since they are constructed from the same set of bands in the same energy window. The key difference lies in the way the interactions are included on the DMFT level: A local interaction on the bond-centered orbitals corresponds to both on-site and inter-site interactions within the atom-centered basis.

For brevity, we refer to the bond-centered orbitals formed from the $a_{1g}$ (or $e_{g}^\pi$) atom-centered orbitals simply as the $a_{1g}$ (or $e_{g}^\pi$) orbitals. Selected constructed bond-centered orbitals are shown in Fig.~\ref{fig:bc-wfs}. In particular, we show the key $a_{1g}$ Wannier orbital for both SB and LB V--V pairs in the M1 structure~[Fig.~\ref{fig:bc-wfs}(a, b)], and for the two identical bonds (B1 and B2) in the R structure~[Fig.~\ref{fig:bc-wfs}(c, d)]. In Fig.~\ref{fig:bc-wfs}(e-h), we also show the $e_{g}^{\pi}$ orbitals for the R structure, which complete the orbital basis set used in our DFT+DMFT calculations, emphasizing that we treat all orbitals within the bond-centered approach. 

\subsection{Computational procedure}

Next, we describe how we implement the DFT+DMFT calculations in the bond-centered basis. For all calculations, we use a monoclinic cell constructed from the lattice vectors of the experimental R structure~\cite{McWhan_et_al:1974} doubled along the $c$ direction (see Fig.~\ref{fig:bc-wfs}). In our treatment of the M1 phase, for simplicity, we disregard the experimentally observed unit cell expansion along $c$ and also all other strain components relative to the R phase~\cite{Longo_et_al:1970}.  We have verified that this yields only very small changes in the results. Embedding the R and M1 structure in the same unit cell and using fixed lattice vectors simplifies the interpolation of atomic positions between the two cases, and allows us to systematically monitor the change in properties as function of the internal structural distortion.

We perform DFT calculations using the \textsc{Quantum ESPRESSO}~(v7.0) package~\cite{Giannozzi_et_al:2009, Giannozzi_et_al:2017} within the generalized gradient approximation using the Perdew-Burke-Ernzerhof~\cite{Perdew/Burke/Ernzerhof:1996} exchange-correlation functional. We use the scalar-relativistic ultrasoft pseudopotentials from the GBRV library~\cite{Garrity_et_al:2014} with semicore 3$s$ and 3$p$ states included as valence for the V~atoms. For good convergence, we use a wavefunction plane-wave kinetic energy cut-off of 70~Ry ($\approx$ 816\,eV) and a 12-times higher cut-off for the charge density. We converge the total energies to 10$^{-8}$\,eV and use a $\Gamma$-centered $6 \times 6 \times 8$ \textbf{k}-point grid.

To construct Wannier functions, we use \textsc{wannier90}~(v3.1.0)~\cite{Mostofi_et_al:2014, Pizzi_et_al:2020}. We use a frozen energy window from $-0.5$ to 1.7\,eV around the Fermi level, which encompasses most of the $t_{2g}$-dominated band manifold. These bands are not entangled with the O~2$p$-dominated bands below, and only very weakly with the higher-energy V~$e_g$-dominated bands [see the density of states (DOS) in Fig.~\ref{fig:bs-plot}~(b)]. For disentanglement, we hence include these higher-lying bands until around 4\,eV. As seen in Fig.~\ref{fig:bs-plot}~(a), the resulting Wannier bands closely reproduce the DFT band structure in the relevant energy range. We work with the Wannier functions obtained directly after the disentanglement without any further spread minimization. 

We then perform one-shot as well as charge self-consistent DFT+DMFT calculations in the bond-centered basis using \texttt{solid\_dmft}~\cite{Merkel_et_al:2022, Beck_et_al:2022} within the TRIQS~(v3.1.0) package~\cite{Parcollet_et_al:2015}. We solve the DMFT impurity problem with the continuous-time quantum Monte-Carlo solver CT-HYB~\cite{Seth_et_al:2016} at the inverse electronic temperature $\beta = (k_B T)^{-1} = 40$ eV$^{-1}$, corresponding to approximately room temperature. We use 10$^4$ warm-up cycles and 4$\times$10$^8$ Monte-Carlo cycles with 120 steps each. We represent the Green's functions using 30 Legendre coefficients~\cite{Boehnke_et_al:2011}, which leads to a smooth self-energy on the imaginary-frequency axis. We average over both spin channels to ensure a paramagnetic solution.

To allow for symmetry lowering relative to the R phase, we treat each bond center as an independent impurity problem, even in the R phase itself, where both bond centers are in principle identical. For one-shot calculations, we converge the orbital and site occupations to 10$^{-2}$ electrons. For charge self-consistent calculations, we first converge the DMFT observables to the same standard as in the one-shot case, and then consecutively perform one DFT iteration with updated charge density followed by one DMFT iteration, until the total energy is converged to $5 \times 10^{-2}$\,eV. We give the final values averaged over the last 10 iterations. We use a Hubbard-Kanamori Hamiltonian including spin-flip and pair-hopping terms (see, e.g., \cite{Georges/Medici/Mravlje:2013}) to represent the local interaction within the bond-centered orbitals on each bond center; this will be justified {\it a posteriori} in Sec.~\ref{subsec:crpa-res}. The DFT+DMFT framework necessitates a double-counting correction, and we use the so-called ``fully localized limit'', \mbox{$\Sigma_{\text{DC}} = (U-2J)(n-1/2)$}~\cite{Held:2007}, which depends on the total site occupation $n$. For consistency with the charge self-consistent calculations, DMFT occupations are used to evaluate the double counting correction in our one-shot calculations.

From the local Green's function, $G_{mm'}(\tau)$, where $m, m'$ label different orbitals and $\tau$ is imaginary time, we obtain the local occupations on a given site (bond center), $n_{mm'}=G_{mm'}(\tau=0^{-})$, as well as the averaged spectral weight around the Fermi level, \mbox{$\bar{A}(\omega=0)=-(\beta/\pi)$Tr$G(\tau=\beta/2)$}. We also calculate the orbital-specific quasiparticle weight from the imaginary part of the local self-energy, by fitting with a third-order polynomial in the lowest five Matsubara frequencies and then interpolating $Z_m = [1-\partial$Im$\Sigma_m(i\omega)/\partial(i\omega)]^{-1}$ to $i\omega=0$. Finally, we use the maximum-entropy method~\cite{Jarrell/Gubernatis:1996, Kraberger_et_al:2017} to obtain the \textbf{k}-averaged spectral functions on the real frequency axis.

\subsection{cRPA calculations}

We calculate the screened Coulomb interaction at zero frequency using the constrained random-phase approximation (cRPA)~\cite{Aryasetiawan_et_al:2004, Miyake/Aryasetiawan:2008} both for the atom-centered and bond-centered basis. Within cRPA, screening is treated within the random phase approximation and is divided into contributions stemming from electronic transitions completely within the correlated subspace, between the correlated subspace and all other bands, and completely within the other bands. The last two contributions then define the frequency-dependent partially screened interaction experienced by the electrons within the correlated subspace. Matrix elements of the screened interaction are then evaluated with the specific Wannier basis representing the correlated subspace. 

We conduct the cRPA calculations using the \textsc{respack} package~\cite{Nakamura_et_al:2021} and the interface tool \texttt{wan2respack}~\cite{Kurita_et_al:2023} to use our custom Wannier functions. For the DFT step, we use an identical $\textbf{k}$-point mesh and energy cutoff as outlined above and find we obtain good convergence if we include 150 empty bands and set the polarization function cut-off to 10~Ry.

We extract the reduced interaction matrices \mbox{$U^{\sigma \sigma}_{mm'}$ = $U_{mm'mm'}$} and \mbox{$U^{\sigma \bar{\sigma}}_{mm'}$ = $U_{mm'mm'} - U_{mm'm'm}$} corresponding to equal and opposite spins, respectively, for further analysis. Finally, we evaluate averaged parameters corresponding to a simple Hubbard-Kanamori parameterization directly from the reduced interaction matrices, where for three orbitals ($N=3$)~\cite{Vaugier/Jiang/Biermann:2012}:
\begin{equation} 
\label{eq_uj_avg}
\begin{split}
    U & = \frac{1}{N} \sum_{m=1}^{N=3} U_{mmmm}, \\
    U'& = \frac{1}{N(N-1)} \sum_{m \neq m' = 1}^{N=3} U_{mm'mm'}, \\
    J & = \frac{1}{N(N-1)} \sum_{m \neq m' = 1}^{N=3} U_{mm'm'm}.
\end{split}
\end{equation}
The validity of the Hubbard-Kanamori parameterization for the bond-centered orbital basis is discussed in Sec~\ref{subsec:crpa-res}.


\section{Results and Discussion}

\subsection{cRPA results}
\label{subsec:crpa-res}

\begin{table}
\caption{The Hubbard-Kanamori screened and unscreened parameter values in eV as obtained from cRPA for the bond-centered (BC) and atom-centered (AC) basis for the R phase.}
\begin{tabular}{p{1.7cm}p{2cm}|R{1.3cm}R{1.3cm}R{1.3cm}}
\hline
\hline
&      (all in eV) & $U$ & $U'$ & $J$ \\ \hline
\multirow{3}{*}{\begin{tabular}[c]{@{}l@{}} AC basis \end{tabular}} & screened &  3.02 &  1.93 &  0.55 \\
                                                                    & unscreened   & 15.38 & 14.12 &  0.63 \\
                                                                    & ratio      &  0.20 &  0.14 &  0.86 \\[0.4cm]
\multirow{3}{*}{\begin{tabular}[c]{@{}l@{}} BC basis \end{tabular}} & screened &  1.35 &  0.97 &  0.19 \\
                                                                    & unscreened   &  7.64 &  7.21 &  0.22 \\
                                                                    & ratio      &  0.18 &  0.14 &  0.86  \\ 
\hline
\hline
\end{tabular}
\label{table_crpa}
\end{table}

We first present and analyze the results of the cRPA calculations for both the atom- and bond-centered basis, to better estimate the strength of the electron-electron interactions within this unconventional basis set and to motivate the parameterization of the local interaction used in our subsequent DFT+DMFT calculations. 

Our calculated reduced screened interaction matrices for the atom-centered (AC) basis in the R structure are [ordered as $a_{1g}$, $e_g^\pi(1)$, and $e_g^\pi(2)$]:
\begin{equation} 
\label{eq_apar_ac_r_crpa}
U^{\sigma \bar{\sigma}}_{\text{AC}} = 
\begin{pmatrix}
3.04 & 1.91 & 1.94 \\
1.91 & 2.91 & 1.93 \\
1.94 & 1.93 & 3.11 \\
\end{pmatrix} \text{eV},
\end{equation}
\begin{equation} 
\label{eq_par_ac_r_crpa}
U^{\sigma \sigma}_{\text{AC}} = 
\begin{pmatrix}
0.00 & 1.39 & 1.38 \\
1.39 & 0.00 & 1.37 \\
1.38 & 1.37 & 0.00 \\
\end{pmatrix} \text{eV}.
\end{equation}
The averaged Hubbard-Kanamori parameters $U$, $U'$, and $J$ are listed in Table~\ref{table_crpa}, together with the corresponding unscreened values. One can see that the different interorbital elements of both $U^{\sigma \sigma'}_{\text{AC}}$ and $U^{\sigma \sigma}_{\text{AC}}$ show almost no variation and are well described by the averaged $U'$ and $U'-J$. The intra-orbital interactions in $U^{\sigma \sigma'}_{\text{AC}}$ show slightly more variation, with deviations of up to 0.11\,eV from the averaged $U$, but overall the screened interaction matrices obtained from cRPA are well approximated by the simplified Hubbard-Kanamori form. 

The Hubbard parameter $U$ is strongly screened to 20~\% of its unscreend value, while  $J$ is only weakly screened to 86~\% of the unscreened value~(see Table~\ref{table_crpa}). We thus obtain a moderate value of $U=3.02$\,eV. This is consistent with previous studies working in a similar atom-centered basis, where $U=3.5-4.2$\,eV was used, leading to a correct assigment of the R metallic and M1 singlet-insulating regimes~\cite{Liebsch/Ishida/Bihlmayer:2005, Biermann_et_al:2005, Kim_et_al:2022}. cRPA calculations for VO$_2$ by Shih~{\it et al.}~\cite{Shih_et_al:2012} used a different definition for the orbitals and the correlated subspace, preventing a direct comparison. We also note that we obtain almost identical values for the parent R phase as for the distorted and dimerized M1 phase~(see Appendix~\ref{app_crpa-res}). 

We now present the screened interaction matrices for the R phase in the bond-centered (BC) basis:
\begin{equation} 
\label{eq_apar_bc_r_crpa}
U^{\sigma \bar{\sigma}}_{\text{BC}} = 
\begin{pmatrix}
1.38 & 0.98 & 0.99 \\
0.98 & 1.30 & 0.96 \\
0.99 & 0.96 & 1.36 \\
\end{pmatrix} \text{eV},
\end{equation}
\begin{equation} 
\label{eq_par_bc_r_crpa}
U^{\sigma \sigma}_{\text{BC}} = 
\begin{pmatrix}
0.00 & 0.80 & 0.80 \\
0.80 & 0.00 & 0.78 \\
0.80 & 0.78 & 0.00 \\
\end{pmatrix} \text{eV}.
\end{equation}
Similarly to the atom-centered basis, the inter-orbital off-diagonal elements show only very weak variations, whereas the intra-orbital diagonal elements of $U^{\sigma\bar\sigma}_\text{BC}$ show slightly more variation. The maximum deviation of 0.05\,eV of the intra-orbital elements from the averaged $U$ is even smaller than for the atom-centered basis.
Thus, also for the bond-centered orbitals, the Hubbard-Kanamori parameterization offers a good approximation to the cRPA-calculated interaction matrices. 

The averaged interaction parameters in the bond-centered basis (listed in Table~\ref{table_crpa}) are notably smaller than those of the atom-centered basis. This is clearly due to the reduction of the corresponding unscreened parameters, which are reduced by a factor of two compared to the atom-centered case, consistent with the larger spatial spread of the bond-centered orbitals. 
On the other hand, the screening is almost exactly as strong as for the atom-centered case, with $U$ and $J$ reduced to 18~\% and 86~\%  of the corresponding unscreened values. This is understandable, since the division of the bands into screening and correlated subspaces is identical in the two cases, and only the orbital representation of the correlated subspace differs. 

Surprisingly, we note that for both basis sets, the relation $U'=U-2J$ holds to a good approximation, even though this is in principle only valid in the spherically symmetric case.

Finally, the change of basis presented here leads to large intersite elements in the Coulomb tensor [see Appendix~\ref{app_crpa-res}, Eq.~(\ref{eq_par_bc_m1_crpa})], where, in particular, the nearest-neighbor interaction terms are similar in magnitude to the diagonal blocks of the Coulomb matrix captured by the Hubbard-Kanamori approximation. Since one of the key ideas of this work is to perform computationally relatively undemanding single-site DFT+DMFT calculations, a key approximation that we assume here is that the correlation effects of these intersite elements are weak and thus we only treat them at the DFT level. Based on all of the above, we use a Hubbard-Kanamori parameterization with two parameters $U$ and $J$ (with $U'=U-2J$) to represent the local Coulomb interaction between the bond-centered orbitals within our DFT+DMFT calculation. 

\begin{figure}
	\centering
	\includegraphics[width=1\linewidth]{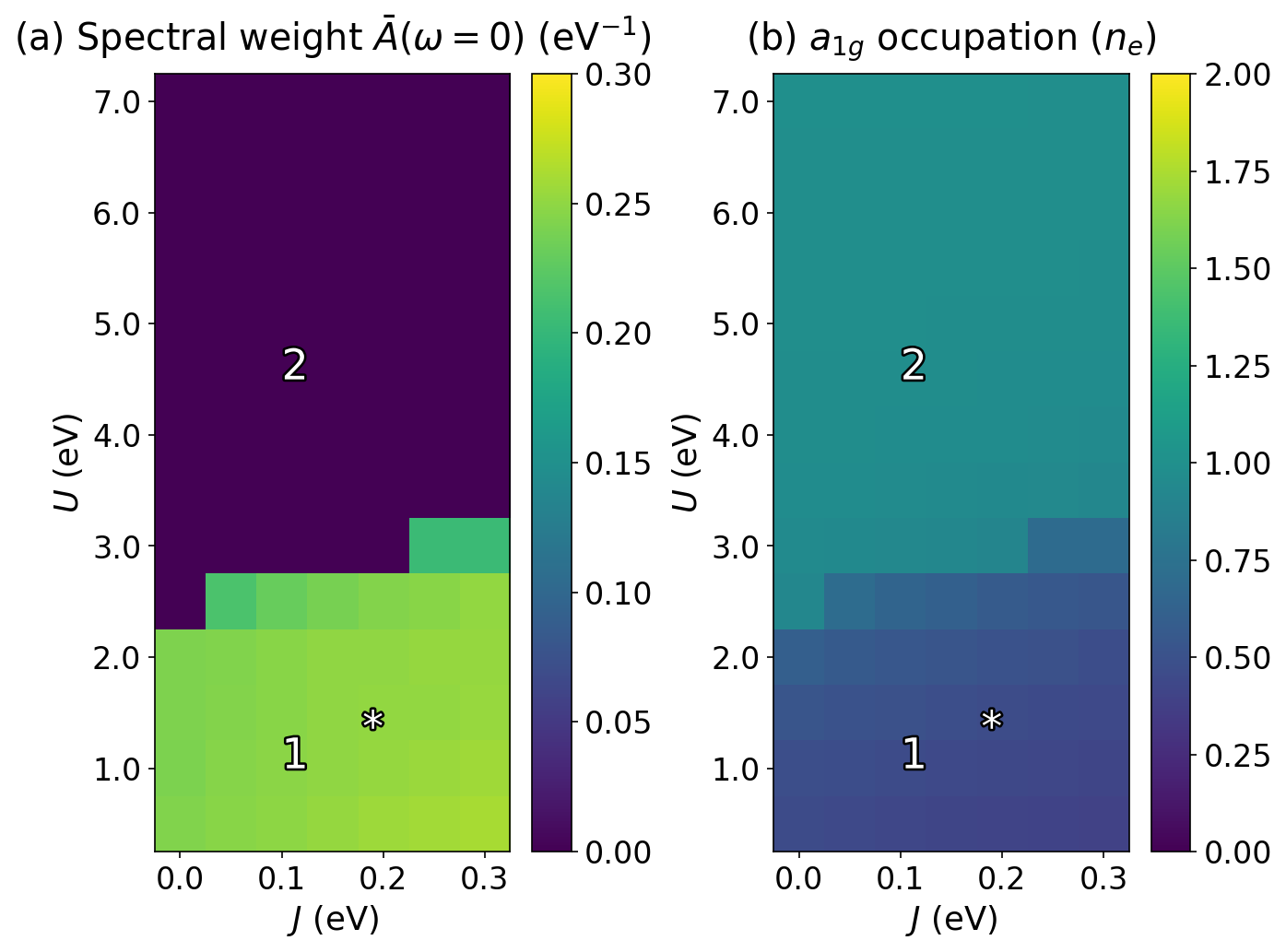}
	\caption{The key observables of one of the identical R bond sites within the DFT+DMFT method as a function of $U$ and $J$. A star indicates cRPA values and the labels (1, 2) correspond to the different regimes as described in the text. (a)~Spectral weight at zero frequency. (b)~Occupation of the $a_{1g}$ orbital in number of electrons, $n_e$.}
	\label{fig:uj-diag-rut}
\end{figure}

\subsection{Phase diagram in \textit{U} and \textit{J}}

We now perform simple one-shot DFT+DMFT calculations for a wide range of interaction parameters to explore the overall behavior of the bond-centered description, which is the only basis we consider from now on. In particular, we present selected observables as a function of the Hubbard-Kanamori model parameters $U$ and $J$. We have verified, at least for the most relevant regime, that charge self-consistency does not lead to substantial changes in the results. To facilitate a cleaner comparison with the charge self-consistent results, we calculate the double-counting correction using DMFT occupations.

\begin{figure*}[!t]
    \centering
    \begin{minipage}[t]{0.48\textwidth}
	\centering
	\includegraphics[width=1\linewidth]{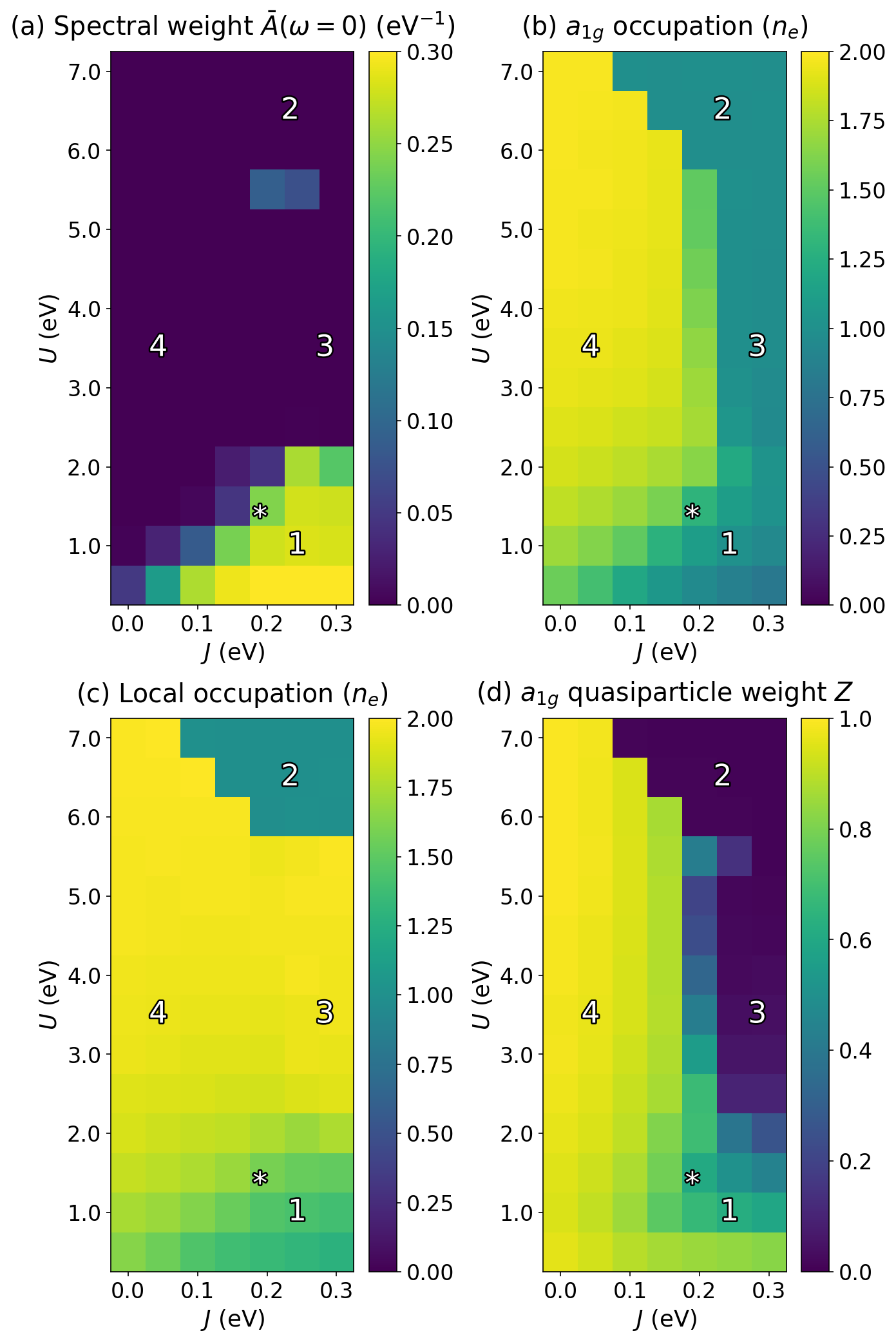}
	\caption{The key observables of the M1 short-bond site obtained within the bond-centered DFT+DMFT method as a function of $U$ and $J$. A star indicates the cRPA values of $U$ and $J$ and the labels (1-4) correspond to the different regimes as described in the text.  (a)~Spectral weight at zero frequency. (b)~Occupation of the $a_{1g}$ orbital in number of electrons, $n_e$. (c)~Local occupation in number of electrons, $n_e$. (d)~The $a_{1g}$ orbital quasiparticle weight.}
	\label{fig:uj-diag}
    \end{minipage}\hfill
    \begin{minipage}[t]{0.48\textwidth}
	\centering
	\includegraphics[width=1\linewidth]{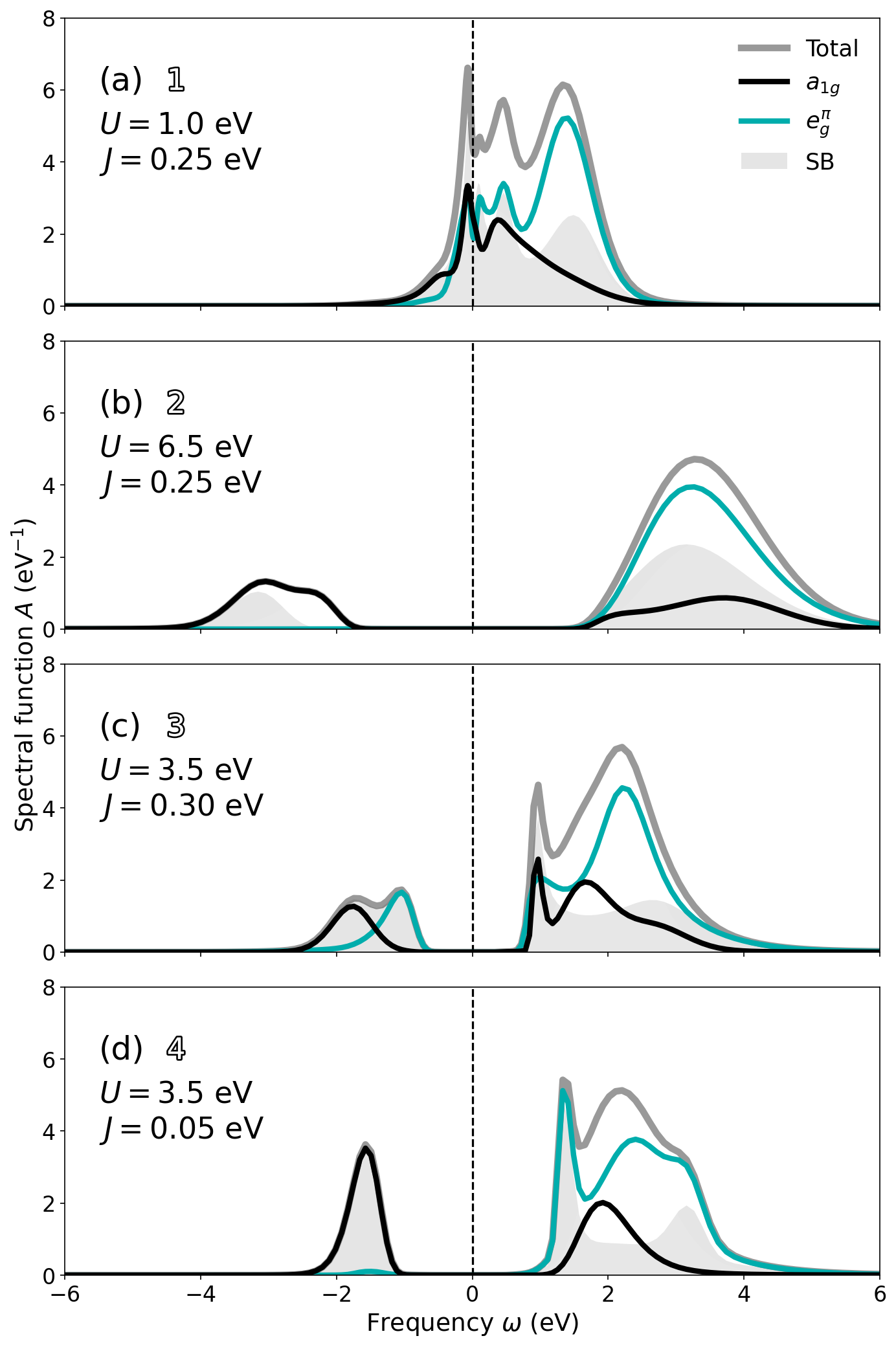}
	\caption{Total spectral functions summed over both SB and LB sites for $a_{1g}$ ($e_{g}^{\pi}$) orbitals shown in black (light-blue) lines. The gray line indicates the sum over all orbitals and area shaded in gray corresponds to the SB site. The figure illustrates specific points corresponding to the different regimes of Fig.~\ref{fig:uj-diag}. (a)~Regime~1, a metal. (b)~Regime~2, a conventional Mott insulator. (c)~Regime~3, a charge-disproportionated Mott insulator. (d)~Regime~4, a singlet insulator.}
	\label{fig:uj-spec}
    \end{minipage}
\end{figure*}

We first discuss the R phase within the bond-centered description, where we always obtain identical results for both bond-centers, due to the equidistant V atoms. We find that the R phase hosts two different regimes, which we illustrate in Fig.~\ref{fig:uj-diag-rut}(a, b) with two observables - the spectral weight around zero frequency and the $a_{1g}$ orbital occupation. In regime 1, at small $U$, we observe a metallic regime [$\bar{A}(0) > 0$] with fractional occupation of all orbitals. Regime 2 occurs at $U \gtrsim 3$\,eV and corresponds to a conventional Mott insulator with a single electron in each of the lowest-lying $a_{1g}$ bond-centered orbitals. As indicated by the star in Fig.~\ref{fig:uj-diag-rut}(a, b), the cRPA values lie deep in the metallic regime. We note that we obtain identical regimes but for different values of $U$ within atom-centered DFT+DMFT, validating the applicability of our bond-centered approach for the R phase.

In the distorted M1 structure, we have inequivalent SB and LB sites. Since the main physics occurs on the short V--V bonds, we mostly discuss the SB observables, referring to the LB sites or global observables only where necessary. For the M1 phase, we observe a richer phase diagram than for the R structure, noting four distinct regimes in the relevant part of the $U$-$J$ phase diagram, denoted 1-4 in Fig.~\ref{fig:uj-diag}. Here, in addition to the spectral weight at zero frequency and the SB $a_{1g}$ orbital occupation, we plot the total occupation on the SB site and the corresponding $a_{1g}$ orbital quasiparticle weight $Z$, representative of the behavior of the self energy around zero frequency. To further characterize the different regimes, we also depict representative spectral functions for regimes 1-4 in Fig.~\ref{fig:uj-spec}(a-d), including also information about the LB site.

In regime~1, at low values of $U$ (and also depending on $J$), we observe a metallic state~[Fig.~\ref{fig:uj-diag}(a)] in which the electron-electron interaction is not sufficiently large to open a Mott-like gap. Although the detailed occupancies of the different orbitals, and also the distribution between SB and LB sites, differ depending on the precise $U$ and $J$ values within this regime, Fig.~\ref{fig:uj-spec}(a) shows that all orbitals on both SB and LB sites contribute to the metallicity.

In regime~2, at relatively large values of $U$ and not too small $J$, we observe a typical Mott-insulating state, analogous to that in the R phase, with identical SB and LB occupations of exactly one electron each [Fig.~\ref{fig:uj-diag}(c)]. Within this regime, the single electron on each site resides in the lowest-lying $a_{1g}$ orbital, leading to vanishing spectral and quasiparticle weights [Fig.~\ref{fig:uj-diag}(a) and (d)]. The spectral function in this regime shows one broad feature consisting of two lower Hubbard peaks below the Fermi level, both of the $a_{1g}$ type from the two respective sites [the lower Hubbard peak belonging to the SB site is shown in gray in Fig.~\ref{fig:uj-spec}(b)].

Lowering the $U$ value for $J \gtrsim 0.2$\,eV favors the formation of an unconventional Mott insulator in regime~3, where we observe a charge-disproportionated Mott state with a doubly-occupied SB site [Fig.~\ref{fig:uj-diag}(c)] with one electron in the $a_{1g}$ [Fig.~\ref{fig:uj-diag}(b)] and one electron in the next-lowest-lying orbital, $e_{g}^{\pi}(1)$ [shown in Fig.~\ref{fig:uj-spec}(c) as a part of the $e_{g}^{\pi}$]. Similar to regime~2, we see two merged occupied Hubbard peaks in the spectral function, but now both derive from the SB site and have different orbital character. In both of these Mott-like states (regimes 2 and 3), we observe a vanishing $a_{1g}$ orbital quasiparticle weight~[Fig.~\ref{fig:uj-diag}(d)]. While in regime~2, the $a_{1g}$ orbital quasiparticle weight is zero for both sites, in regime~3 this is only true for the SB site. We also note a metallic boundary between regime~2 and regime~3, as seen in Fig.~\ref{fig:uj-diag}(a) for two data points with non-zero spectral weight, persisting to larger values in $J$.

Finally, in regime~4, we observe an insulating state where the SB site hosts two electrons [Fig.~\ref{fig:uj-diag}(c)], both within the $a_{1g}$ bonding orbital [Fig.~\ref{fig:uj-diag}(b)]. Despite the insulating character, the self-energy of this state remains small and hence the system still retains a finite $a_{1g}$ orbital quasiparticle weight $Z \approx 1$. The finite quasiparticle weight can also be traced to the relatively sharp quasiparticle-like feature in the occupied part of the spectral function, shown in Fig.~\ref{fig:uj-spec}(d). This rather weakly correlated regime is the singlet insulator state which corresponds to the experimentally observed M1 ground state. At small values of $J$, this region spans a wide range of $U$ values which is likely due to the almost insulating nature of the DFT calculations~[see the dip in the DOS at the Fermi level in Fig.~\ref{fig:bs-plot}(a)]. We also note that close to the boundary to the metallic state, the occupations change gradually, i.e., the $a_{1g}$ and local occupations are less than two. The shape of the spectral function in this region [Fig.~\ref{fig:uj-spec}(d)] is in good agreement with previous works simulating the M1 structure of VO$_2$ with more expensive cluster-DMFT methods such as those from Refs.~\cite{Biermann_et_al:2005, Belozerov_et_al:2012}.

Our bond-centered method thus offers a substantial realistic $U$ and $J$ parameter range for which we are able to describe a metallic R phase and an insulating M1 phase. This region spans essentially the whole singlet insulator regime of M1 below $U \lesssim 3$\,eV, above which the R phase changes into a Mott insulator. Out of the regimes discussed, the cRPA calculations from Section~\ref{subsec:crpa-res}~[denoted by stars in Fig.~\ref{fig:uj-diag-rut}(a, b) and Fig.~\ref{fig:uj-diag}(a-d)] fall directly into the metallic regime in the R phase and on the boundary of the metallic and singlet-insulator regimes in M1. However, we note that cRPA values should merely be taken as a guideline since the frequency dependence of the screened interaction is neglected in the DFT+DMFT calculations with only the zero frequency component used, and since the cRPA method is known to overscreen in some cases~\cite{Honerkamp_et_al:2018}.

\begin{figure*}[!t]
    \centering
    \begin{minipage}[t]{0.48\textwidth}
	\centering
	\includegraphics[width=1\linewidth]{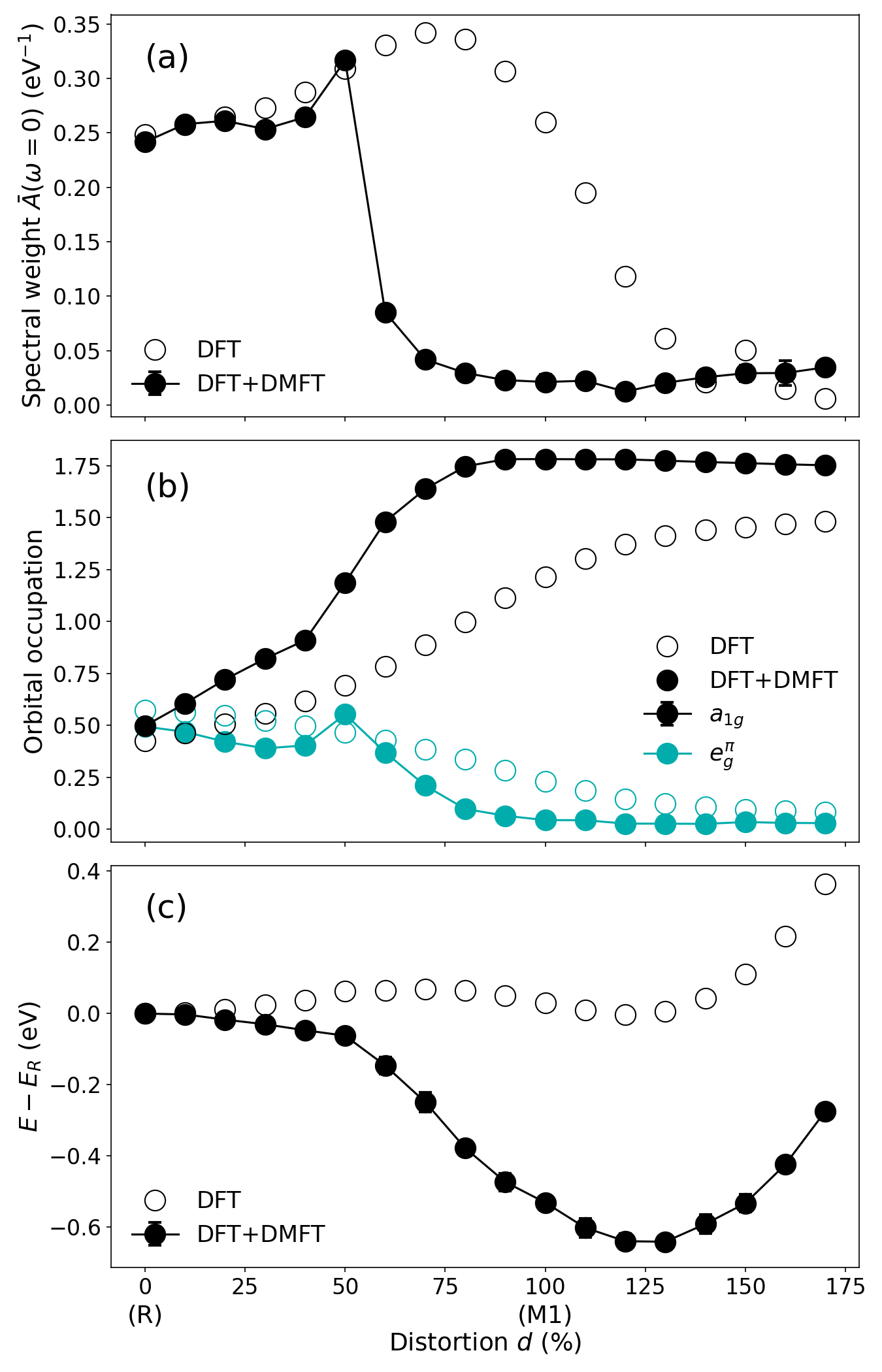}
	\caption{Selected DFT(+DMFT) observables in empty (filled) markers along the distortion interpolating between and extrapolating beyond the R and M1 VO$_2$ phases. (a)~Short-bond spectral weight at zero frequency. (b)~Short-bond occupation of the $a_{1g}$ ($e_{g}^{\pi}$) orbitals in black (light-blue). (c)~Total energy relative to the R phase.}
	\label{fig:dist-obs}
    \end{minipage}\hfill
    \begin{minipage}[t]{0.48\textwidth}
	\centering
	\includegraphics[width=1\linewidth]{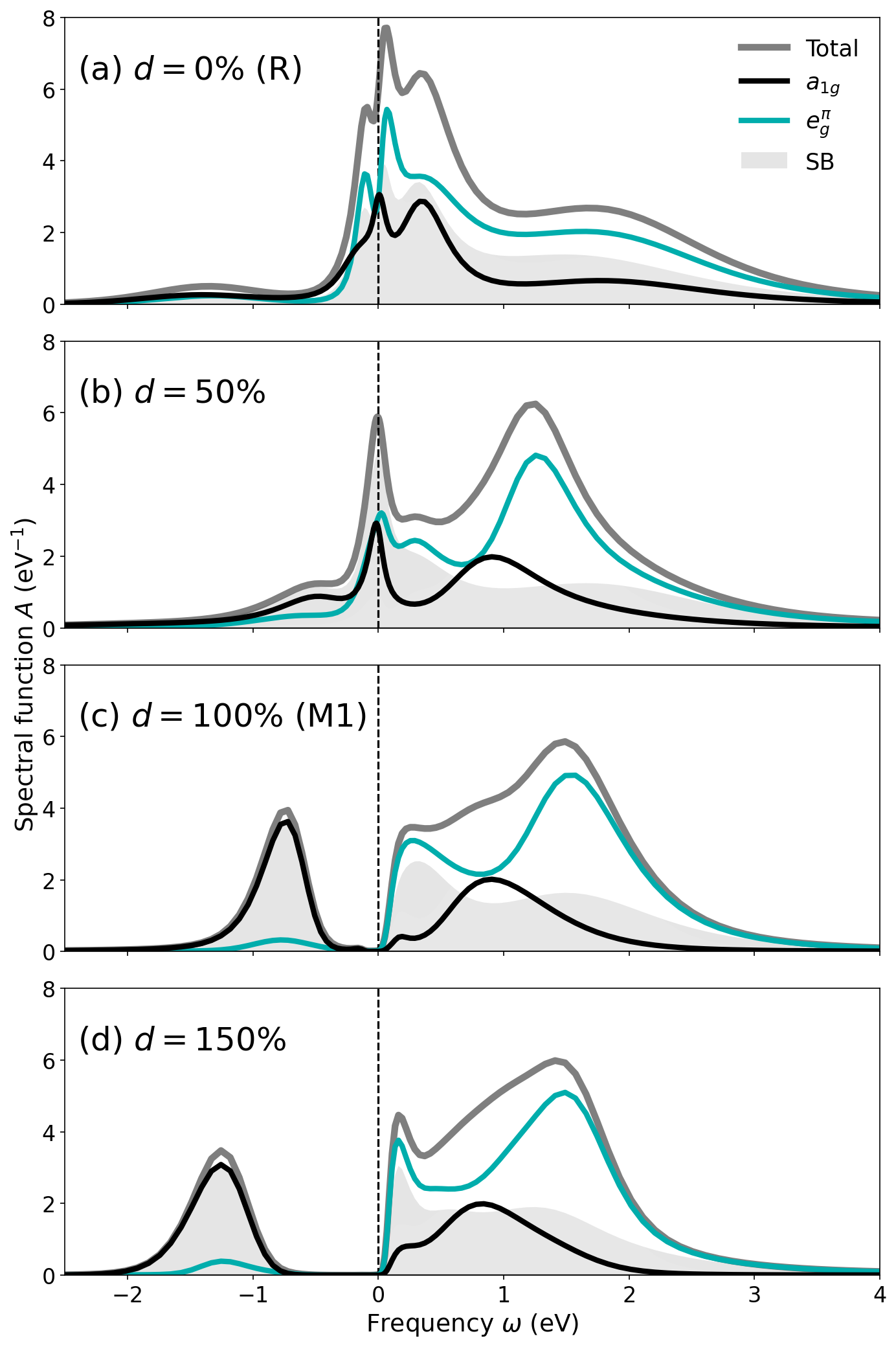}
	\caption{
 Total spectral functions at four specified distortions, $d$, summed over both SB and LB sites for $a_{1g}$ ($e_{g}^{\pi}$) orbitals shown in black (light-blue) lines. The gray line indicates the sum over all orbitals and the area shaded in gray corresponds to the SB site.}
	\label{fig:dist-spec}
    \end{minipage}
\end{figure*}

\subsection{Variation of the M1 structural distortion}
\label{sec:dist}

Having identified the regime within our $U$ and $J$ phase diagrams that matches the experimental observations, we now perform a variation of the structural distortion in this regime to analyze the onset of the MIT with distortion. Taking into account the limitations of the cRPA approach and the fact that our DMFT calculation does not account for any frequency dependence of the screening, we choose $U=2.0$\,eV and $J=0.1$\,eV for this study. These values are clearly within the singlet insulator regime for the M1 structure and still within the metallic regime for the R case. We linearly interpolate between and extrapolate beyond the atomic positions of the R (corresponding to $d=0\,\%$) and M1 (corresponding to $d=100\,\%$) structures with fixed lattice vectors. We perform charge self-consistent DFT+DMFT so that we also have access to the total energy of the system. In Fig.~\ref{fig:dist-obs}(a-c), we present the SB (or one of the bond sites for R) spectral weight at zero frequency, SB occupations, and the total energy as function of the structural distortion, together with the respective DFT counterparts (empty circles in Fig.~\ref{fig:dist-obs}). We do not show data for the LB site (or the other bond site for R) as it quickly depletes its electrons as the distortion increases. Similarly to the previous section, we also show the spectral function plots of both sites in Fig.~\ref{fig:dist-spec}(a-d) at particular points along the distortion path. 

At small distortions away from the R structure, the metallic phase persists [Fig.~\ref{fig:dist-obs}(a)], characterized by the $a_{1g}$ and $e_g^\pi$ orbitals with similar occupations [Fig.~\ref{fig:dist-obs}(b)]. At around $d=50\,\%$ distortion, we start seeing the electrons strongly localizing in the $a_{1g}$ orbital of the SB, with the $e_{g}^{\pi}$ spectral weight shifting to higher energies [Fig.~\ref{fig:dist-spec}(b)]. In this regime close to the R structure, DFT gives similar results, with a slightly lower population of the $a_{1g}$ orbital compared to the DFT+DMFT case.

Between around $d=50\,\%$ and $d=70\,\%$, in the DFT+DMFT results, we observe an abrupt change in both the spectral weight at the Fermi level and the occupations of the $a_{1g}$ orbital, corresponding to a change from a metal to the singlet insulator. We also see a clear change in the curvature in the total energy~[Fig.~\ref{fig:dist-obs}(c)]. These sharp changes are in stark contrast to the DFT results on multiple fronts. Firstly, DFT+DMFT predicts a sharp MIT at $d=70\,\%$ [Fig.~\ref{fig:dist-obs}(a)], whereas DFT predicts that the system stays metallic until a large distortion of around $d=120\,\%$. Secondly, the very large increase of the $a_{1g}$ SB occupation at the MIT predicted by DFT+DMFT [Fig.~\ref{fig:dist-obs}(b)] is not present in the DFT results. Finally, while DFT predicts two energy minima, one for the undistorted structure and another slightly higher-energy one at around 125\,\% distortion [Fig.~\ref{fig:dist-obs}(c)], we obtain only a single energy minimum for the distorted case within DFT+DMFT. 

Above $d=70\,\%$, DFT+DMFT retains the singlet insulating state and we see a shift of the bonding $a_{1g}$ peak to lower energy, leading to an increasing gap in the spectral function [Fig.~\ref{fig:dist-spec}(c, d)]. At distortions above $d = 120\,\%$, DFT also converges into an insulating state with the SB $a_{1g}$ orbital population saturating at 1.5 electrons. It is further interesting to note that the position of the energy minimum for non-zero distortion does not change significantly with the different methods. Instead, the minimum is considerably deeper in the DMFT case, stabilizing the M1 ground state, consistent with experimental observations. Note, however, that for both DFT and DFT+DMFT the calculated minimum occurs at a larger structural distortion than that of the experimental M1 structure ($d=100\,\%$). Lastly, the spectral functions of both end-point phases (R at $d=0\,\%$ and M1 at $ d=100\,\%$) agree well with previous works~\cite{Biermann_et_al:2005, Belozerov_et_al:2012}.


\section{Summary and Outlook}

In this work, we presented a DFT + single-site DMFT study of VO$_2$ using an unconventional set of bond-centered orbitals to represent the correlated subspace. This allowed us to treat the R and M1 structures in a consistent manner, and thus to distort the system continuously between the two end-point structures. Thus, we can observe and analyze the MIT as a function of distortion at a greatly reduced computational cost compared to cluster-based methods.

We defined a \textbf{k}-dependent transformation matrix that we used to form the bond-centered orbitals from the standard atom-centered ones, and we performed cRPA calculations of the screened electron-electron interaction for both the atom- and bond-centered basis sets. We observed that the bond-centered basis has much smaller values of interaction parameters compared to the atom-centered basis, stemming from its larger spatial spread. 

Performing DFT+DMFT calculations for the bond-centered basis, we then identified key regimes in the $U$ and $J$ phase space for both R and M1 structures, and showed that it is possible to obtain a metallic R phase and singlet-insulator M1 phase consistent with the cRPA results. With the results of the cRPA method, our {\it ab initio} procedure replicates the experimental findings for the different phases. 

Finally, we performed calculations as function of the structural distortion between the R and M1 structures, observing an MIT indicated by a sharp change in spectral weight at the Fermi level and orbital occupations. Furthermore, whereas DFT exhibits a global energy minimum for the R structure, in contradiction with the experimentally verified M1 ground state, DFT+DMFT correctly predicts a global energy minimum for M1.

In conclusion, we find that use of a bond-centered basis set combined with DFT+DMFT offers results complementary to the current state-of-the-art cluster-based methods for VO$_2$ and allows convenient analysis of VO$_2$ as function of the structural distortion. In addition, the bond-centered approach is bias-free, since it, in principle, allows the singlet to form on any of the V--V pairs. While the bond-centered approach is particularly well-suited to describe the physics of VO$_2$, we also expect it to be useful in other materials hosting molecular-orbital-like states that necessitate such a special treatment of interatomic correlations.

\section*{Acknowledgments}
We thank Alberto Carta for useful discussions. This work was supported by ETH Z\"{u}rich and through the Swiss National Science Foundation (Grant No.~209454). Calculations were performed on the ETH Z\"{u}rich Euler cluster and the Swiss National Supercomputing Center Piz Daint and Eiger clusters under Project IDs s1128 and eth3.

\appendix

\section{cRPA results for the M1 structure}
\label{app_crpa-res}

In addition to the results described in Section~\ref{subsec:crpa-res}, here we present the reduced Coulomb matrices obtained within cRPA for the M1 phase. First, we show the results for the atom-centered basis:
\begin{equation} 
\label{eq_apar_ac_m1_crpa}
U^{\sigma \bar{\sigma}}_{\text{AC}} = 
\begin{pmatrix}
3.12 & 1.98 & 1.98 \\
1.98 & 2.93 & 1.99 \\
1.98 & 1.99 & 3.10 \\
\end{pmatrix} \text{eV},
\end{equation}
\begin{equation} 
\label{eq_par_ac_m1_crpa}
U^{\sigma \sigma}_{\text{AC}} = 
\begin{pmatrix}
0.00 & 1.47 & 1.43 \\
1.47 & 0.00 & 1.46 \\
1.43 & 1.46 & 0.00 \\
\end{pmatrix} \text{eV}.
\end{equation}
Here, the matrices still follow the Hubbard-Kanamori form very well with slightly larger orbital dependencies than for the R phase, {\it cf.} Eqs.~(\ref{eq_apar_ac_r_crpa}) and (\ref{eq_par_ac_r_crpa}).

Next, we also present the results for the bond-centered orbitals in the distorted M1 phase, using $6 \times 6$ matrices containing both inequivalent sites. Here, the $3 \times 3$ blocks in the diagonal correspond to the local interactions on the bond-centers, and the off-diagonal blocks indicate the interactions between neighboring bond-centers:
\begin{equation} 
\label{eq_apar_bc_m1_crpa}
U^{\sigma \bar{\sigma}}_{\text{cRPA}} = 
\begin{pmatrix}
\begin{array}{ccc|ccc}
1.46 & 1.03 & 1.04 & 1.02 & 0.75 & 0.75 \\
1.03 & 1.33 & 1.01 & 0.75 & 0.93 & 0.73 \\
1.04 & 1.01 & 1.38 & 0.75 & 0.73 & 0.96 \\
\hline
1.02 & 0.75 & 0.75 & 1.33 & 0.93 & 0.93 \\
0.75 & 0.93 & 0.73 & 0.93 & 1.24 & 0.92 \\
0.75 & 0.73 & 0.96 & 0.93 & 0.92 & 1.30 \\
\end{array}
\end{pmatrix} \text{eV},
\end{equation}
\begin{equation} 
\label{eq_par_bc_m1_crpa}
U^{\sigma \sigma}_{\text{cRPA}} = 
\begin{pmatrix}
\begin{array}{ccc|ccc}
0.00 & 0.85 & 0.85 & 0.50 & 0.64 & 0.64 \\
0.85 & 0.00 & 0.83 & 0.64 & 0.44 & 0.62 \\
0.85 & 0.83 & 0.00 & 0.63 & 0.62 & 0.44 \\
\hline
0.50 & 0.64 & 0.63 & 0.00 & 0.76 & 0.75 \\
0.64 & 0.44 & 0.62 & 0.76 & 0.00 & 0.74 \\
0.64 & 0.62 & 0.44 & 0.75 & 0.74 & 0.00 \\
\end{array}
\end{pmatrix} \text{eV}.
\end{equation}
Again, the Hubbard-Kanamori form is approximately observed for the local interactions, both in the short-bond and long-bond blocks. We also note that the difference between the two bond centers is relatively small, and comparable to the orbital dependencies in the R phase, {\it cf.} Eqs.~(\ref{eq_par_bc_r_crpa}) and (\ref{eq_apar_bc_r_crpa}). Overall, the differences in the local interaction parameters between the M1 and R phases are not significant for the comparison between the DFT+DMFT results for the two cases, and the use of constant $U$ and $J$ values for the variation of the structural distortion in Sec.~\ref{sec:dist} appears justified.
Finally, we note the large intersite elements which exhibit similar orbital dependence as the onsite terms.

\section{Centers of the transformed Wannier functions}
\label{app_rot-mat}

Here, we show that the pairwise unitary transformation defined by Eq.~(\ref{eq_rotmat}) indeed leads to Wannier functions that are centered on the mid-point between two consecutive V atoms along the $c$ direction.

The unitary transformation acting in $\textbf{k}$-space was defined as:
\begin{equation} 
U(\textbf{k}) = \frac{1}{\sqrt{2}}
\begin{pmatrix}
 e^{i(\frac{\pi}{4} - \frac{c}{4} k_z)} & e^{i(-\frac{\pi}{4} + \frac{c}{4} k_z)}  \\
 e^{i(-\frac{\pi}{4} - \frac{3c}{4} k_z)} & e^{i(\frac{\pi}{4} - \frac{c}{4} k_z)}  \\
\end{pmatrix},
\end{equation}
where $c$ is the lattice parameter of the monoclinic structure along the $\mathbf{\hat{z}}$ direction, i.e. along the dimerization direction, and $U(\mathbf{k})$ always acts on a pair of $\mathbf{k}$-space Wannier functions corresponding to atom-centered Wannier functions with the same orbital character on both atoms, $|w_{1, \mathbf{k}} \rangle$ and $|w_{2, \mathbf{k}} \rangle$. The transformation into the ``bond-centered'' $\mathbf{k}$-space Wannier orbitals, $|\widetilde{w}_{1,\mathbf{k}} \rangle$ and $|\widetilde{w}_{2,\mathbf{k}} \rangle$ can thus be written as
\begin{equation}
    \begin{pmatrix}
     | \widetilde{w}_{1, \mathbf{k}} \rangle \\
     | \widetilde{w}_{2, \mathbf{k}} \rangle
    \end{pmatrix}
    = U(\mathbf{k})  
    \begin{pmatrix}
     | w_{1, \mathbf{k}} \rangle \\
     | w_{2, \mathbf{k}} \rangle
    \end{pmatrix}.
\end{equation}
The corresponding real-space Wannier functions are then obtained, as usual, by a simple Fourier transform~\cite{Marzari_et_al:2012}
\begin{equation}
| \widetilde{w}_{n, \mathbf{R}} \rangle = \frac{V}{(2\pi)^3} \int d\mathbf{k} \, e^{-i \mathbf{k} \cdot \mathbf{R}} | \widetilde{w}_{n, \mathbf{k}} \rangle ,
\end{equation}
where $V$ is the unit cell volume.

For convenience, we also define two additional sets of partially transformed Wannier functions
\begin{align}
\label{eq_wprimes}
| w_{1, \mathbf{R}}^\prime \rangle &= \frac{V}{(2\pi)^3} \int d\mathbf{k} \, e^{-i\mathbf{k}\mathbf{R} - i k_z \frac{c}{4}} | w_{1, \mathbf{k}} \rangle, \\
| w_{2, \mathbf{R}}^\prime \rangle &= \frac{V}{(2\pi)^3} \int d\mathbf{k} \, e^{-i\mathbf{k}\mathbf{R}  + i k_z \frac{c}{4}} | w_{2, \mathbf{k}} \rangle ,
\end{align}
and
\begin{align}
| w_{1, \mathbf{R}}^{\prime\prime} \rangle &= \frac{V}{(2\pi)^3} \int d\mathbf{k} \, e^{-i\mathbf{k}\mathbf{R}  - i k_z \frac{3c}{4}} | w_{1, \mathbf{k}} \rangle, \\
| w_{2, \mathbf{R}}^{\prime\prime} \rangle &= \frac{V}{(2\pi)^3} \int d\mathbf{k} \, e^{-i\mathbf{k}\mathbf{R}  - i k_z \frac{c}{4}} | w_{2, \mathbf{k}} \rangle .
\label{eq_wdprimes}
\end{align}
With this, the fully transformed Wannier functions can now be written as
\begin{align}
|\widetilde{w}_{1,\mathbf{R}}\rangle &= e^{i\frac{\pi}{4}} |w_{1,\mathbf{R}}^\prime \rangle + e^{-i\frac{\pi}{4}} |w_{2, \mathbf{R}}^\prime \rangle, \\
|\widetilde{w}_{2,\mathbf{R}} \rangle & = e^{-i\frac{\pi}{4}} |w_{1, \mathbf{R}}^{\prime\prime} \rangle + e^{i\frac{\pi}{4}} |w_{2,\mathbf{R}}^{\prime\prime} \rangle .
\end{align}

For the following, it is important to note that if the original atom-centered Wannier functions, $|w_{n, \mathbf{R}}\rangle$, are real, then the partially transformed functions, $|w_{n,\mathbf{R}}^\prime\rangle$ and  $|w_{n, \mathbf{R}}^{\prime\prime}\rangle$ are also real. 
This can be seen as follows. 
A real valued Wannier function $|w_{n, \mathbf{R}}\rangle$ implies the following condition on the cell-periodic part of the corresponding function in $\mathbf{k}$-space (see Eq.~(65) from Ref.~\cite{Marzari_et_al:2012}): 
\begin{equation}
\label{eq:realWF}
u_{n, \textbf{k}}(\textbf{r}) = u^*_{n, -\textbf{k}}(\textbf{r}) ,
\end{equation}
where $u_{n,\mathbf{k}}(\mathbf{r}) = e^{-i\mathbf{k} \cdot \mathbf{r} } w_{n, \mathbf{k}}(\mathbf{r})$.
Since the partially transformed functions in Eqs.~(\ref{eq_wprimes})-(\ref{eq_wdprimes}) differ from the original functions only by a phase factor with a linear $\mathbf{k}$ dependence, the condition in Eq.~(\ref{eq:realWF}) is retained for the corresponding cell-periodic functions, and thus the partially transformed Wannier functions are also real-valued.

We can now calculate the center of the fully transformed Wannier function $|\widetilde{w}_{1,\mathbf{R}=0}\rangle$:
\begin{equation} 
\begin{split}
\label{eq:center}
\langle \widetilde{w}_{1, \mathbf{R}=0} | \textbf{r} | \widetilde{w}_{1, \mathbf{R}=0} \rangle  = &\langle w_{1, \mathbf{R}=0}^\prime | \textbf{r} | w_{1, \mathbf{R}=0}^\prime \rangle \\
+ &\langle w_{2, \mathbf{R}=0}^\prime | \textbf{r} | w_{2, \mathbf{R}=0}^\prime \rangle \\
+ e^{-i\frac{\pi}{2}} &\langle w_{1, \mathbf{R}=0}^\prime | \textbf{r} | w_{2, \mathbf{R}=0}^\prime \rangle \\
+ e^{+i\frac{\pi}{2}} &\langle w_{2, \mathbf{R}=0}^\prime | \textbf{r} | w_{1, \mathbf{R}=0}^\prime \rangle .
\end{split}
\end{equation}
Using the fact that the functions $| w_{n, \mathbf{R}=0}^\prime \rangle$ are real and that $\textbf{r}$ is a Hermitian operator, we can see that the last two terms in Eq.~(\ref{eq:center}) cancel each other:
\begin{equation} 
\begin{split}
& e^{-i\frac{\pi}{2}} \langle w_{1, \mathbf{R}=0}^\prime | \textbf{r} | w_{2, \mathbf{R}=0}^\prime \rangle + e^{+i\frac{\pi}{2}} \langle w_{2, \mathbf{R}=0}^\prime | \textbf{r} | w_{1, \mathbf{R}=0}^\prime \rangle \\
= ( & e^{-i\frac{\pi}{2}} + e^{+i\frac{\pi}{2}}) \langle w_{1, \mathbf{R}=0}^\prime | \textbf{r} | w_{2, \mathbf{R}=0}^\prime \rangle = 0.
\end{split}
\end{equation}
Then, for the diagonal terms, we use the general expression for the expectation value of the position operator for Wannier functions~\cite{Marzari_et_al:2012}:
\begin{equation}
\mathbf{r}_n = \langle w_{n,\mathbf{R}=0} | \mathbf{r} | w_{n, \mathbf{R}=0} \rangle = i \frac{V}{(2\pi)^3} \int d\mathbf{k} \langle u_{n, \mathbf{k}} | \nabla_\mathbf{k} | u_{n, \mathbf{k}} \rangle.
\end{equation}
Thus, a phase transformation of the $\mathbf{k}$-space Wannier functions with a linear $k_z$-dependence simply shifts the center of the corresponding real-space Wannier function according to
\begin{equation}
\begin{split}
\mathbf{r}_n^\prime & =  
i \frac{V}{(2\pi)^3} \int d\mathbf{k} \langle e^{-i k_z \alpha} u_{n, \mathbf{k}} | \nabla_\mathbf{k} | e^{-i k_z \alpha} u_{n, \mathbf{k}} \rangle \\
 &= i \frac{V}{(2\pi)^3} \int d\mathbf{k} [ \langle u_{n, \mathbf{k}} | \nabla_\mathbf{k} | u_{n, \mathbf{k}} \rangle 
 - i \alpha \hat{\textbf{z}} \langle u_{n, \mathbf{k}} | u_{n, \mathbf{k}} \rangle ] \\
 &= \textbf{r}_n + \alpha \hat{\textbf{z}}.
\end{split}
\end{equation}

With this, one can see that the corresponding phase shifts of the first two terms in Eq.~(\ref{eq:center}) cancel each other and one obtains
\begin{equation}
\langle \widetilde{w}_{1, \mathbf{R}=0} | \textbf{r} | \widetilde{w}_{1, \mathbf{R}=0} \rangle =  \frac{1}{2} (\textbf{r}_1 + \textbf{r}_2) ,
\end{equation}
i.e., the center of the first transformed Wannier function indeed lies on the midpoint between the centers of the two original Wannier functions. Thus, if these original Wannier functions are centered on two consecutive V sites, the transformed Wannier function is centered on the corresponding bond center.

An analogous calculation for the second transformed Wannier function $|\widetilde{w}_{2,\mathbf{R}=0}\rangle$ leads to
\begin{align}
\langle \widetilde{w}_{2, \mathbf{R}=0} | \textbf{r} | \widetilde{w}_{2, \mathbf{R}=0} \rangle &= \frac{1}{2} (\textbf{r}_1 + \textbf{r}_2) + \frac{c}{2}\hat{\mathbf{z}} \\
& = \frac{1}{2} ( \left\{ \mathbf{r}_1+c\hat{\mathbf{z}} \right\} + \mathbf{r}_2 ).
\end{align}
Thus, this Wannier function is centered on the next bond center along $\hat{\mathbf{z}}$, i.e., on the midpoint between the second and the periodically shifted first atom-centered Wannier function.  

\bibliography{bc-wfs}

\end{document}